\documentclass{elsart}

\usepackage{graphicx}

\newif\ifpreprint
\preprinttrue

\begin{document}

\journal{Physics Letter B}

\begin{frontmatter}

\date{March 27, 2007}
\title{
\ifpreprint
\rightline{\normalsize FERMILAB--Pub--07/011--E}
\vspace{0.2cm}
\fi
Measurement of the $\Omega_{c}^{0}$ lifetime}


The SELEX Collaboration
\author[Rome]{M.~Iori},
\author[Iowa]{A.S.~Ayan},
\author[Iowa]{U.~Akgun},
\author[PNPI]{G.~Alkhazov},
\author[SLP]{J.~Amaro-Reyes},
\author[PNPI]{A.G.~Atamantchouk\thanksref{tra}},
\author[ITEP]{M.Y.~Balatz\thanksref{tra}},
\author[SLP]{A.~Blanco-Covarrubias},
\author[PNPI]{N.F.~Bondar},
\author[Fermi]{P.S.~Cooper},
\author[Flint]{L.J.~Dauwe},
\author[ITEP]{G.V.~Davidenko},
\author[MPI]{U.~Dersch\thanksref{trb}},
\author[ITEP]{A.G.~Dolgolenko},
\author[ITEP]{G.B.~Dzyubenko},
\author[CMU]{R.~Edelstein},
\author[Paulo]{L.~Emediato},
\author[CBPF]{A.M.F.~Endler},
\author[SLP]{J.~Engelfried\corauthref{cor}},
\corauth[cor]{Corresponding author.}
\ead{jurgen@ifisica.uaslp.mx}
\author[MPI]{I.~Eschrich\thanksref{trc}},
\author[Paulo]{C.O.~Escobar\thanksref{trd}},
\author[ITEP]{A.V.~Evdokimov},
\author[MSU]{I.S.~Filimonov\thanksref{tra}},
\author[Paulo,Fermi]{F.G.~Garcia},
\author[Rome]{M.~Gaspero},
\author[Aviv]{I.~Giller},
\author[PNPI]{V.L.~Golovtsov},
\author[Paulo]{P.~Gouffon},
\author[Bogazici]{E.~G\"ulmez},
\author[Beijing]{He~Kangling},
\author[CMU]{S.Y.~Jun},
\author[Iowa]{M.~Kaya\thanksref{tre}},
\author[Fermi]{J.~Kilmer},
\author[PNPI]{V.T.~Kim},
\author[PNPI]{L.M.~Kochenda},
\author[MPI]{I.~Konorov\thanksref{trf}},
\author[Protvino]{A.P.~Kozhevnikov},
\author[PNPI]{A.G.~Krivshich},
\author[MPI]{H.~Kr\"uger\thanksref{trg}},
\author[ITEP]{M.A.~Kubantsev},
\author[Protvino]{V.P.~Kubarovsky},
\author[CMU,Fermi]{A.I.~Kulyavtsev},
\author[PNPI,Fermi]{N.P.~Kuropatkin},
\author[Protvino]{V.F.~Kurshetsov},
\author[CMU,Protvino]{A.~Kushnirenko},
\author[Fermi]{S.~Kwan},
\author[Fermi]{J.~Lach},
\author[Trieste]{A.~Lamberto},
\author[Protvino]{L.G.~Landsberg\thanksref{tra}},
\author[ITEP]{I.~Larin},
\author[MSU]{E.M.~Leikin},
\author[Beijing]{Li~Yunshan},
\author[UFP]{M.~Luksys},
\author[Paulo]{T.~Lungov},
\author[PNPI]{V.P.~Maleev},
\author[CMU]{D.~Mao\thanksref{trh}},
\author[Beijing]{Mao~Chensheng},
\author[Beijing]{Mao~Zhenlin},
\author[CMU]{P.~Mathew\thanksref{tri}},
\author[CMU]{M.~Mattson},
\author[ITEP]{V.~Matveev},
\author[Iowa]{E.~McCliment},
\author[Aviv]{M.A.~Moinester},
\author[Protvino]{V.V.~Molchanov},
\author[SLP]{A.~Morelos},
\author[Iowa]{K.D.~Nelson\thanksref{trj}},
\author[MSU]{A.V.~Nemitkin},
\author[PNPI]{P.V.~Neoustroev},
\author[Iowa]{C.~Newsom},
\author[ITEP]{A.P.~Nilov\thanksref{tra}},
\author[Protvino]{S.B.~Nurushev},
\author[Aviv]{A.~Ocherashvili\thanksref{trk}},
\author[Iowa]{Y.~Onel},
\author[Iowa]{E.~Ozel},
\author[Iowa]{S.~Ozkorucuklu\thanksref{trl}},
\author[Trieste]{A.~Penzo},
\author[Protvino]{S.V.~Petrenko},
\author[Iowa]{P.~Pogodin\thanksref{trm}},
\author[CMU]{M.~Procario\thanksref{trn}},
\author[ITEP]{V.A.~Prutskoi},
\author[Fermi]{E.~Ramberg},
\author[Trieste]{G.F.~Rappazzo},
\author[PNPI]{B.V.~Razmyslovich\thanksref{tro}},
\author[MSU]{V.I.~Rud},
\author[CMU]{J.~Russ},
\author[Trieste]{P.~Schiavon},
\author[MPI]{J.~Simon\thanksref{trp}},
\author[ITEP]{A.I.~Sitnikov},
\author[Fermi]{D.~Skow},
\author[Bristo]{V.J.~Smith},
\author[Paulo]{M.~Srivastava},
\author[Aviv]{V.~Steiner},
\author[PNPI]{V.~Stepanov\thanksref{tro}},
\author[Fermi]{L.~Stutte},
\author[PNPI]{M.~Svoiski\thanksref{tro}},
\author[PNPI,CMU]{N.K.~Terentyev},
\author[Ball]{G.P.~Thomas},
\author[SLP]{I.~Torres},
\author[PNPI]{L.N.~Uvarov},
\author[Protvino]{A.N.~Vasiliev},
\author[Protvino]{D.V.~Vavilov},
\author[SLP]{E.~V\'azquez-J\'auregui},
\author[ITEP]{V.S.~Verebryusov},
\author[Protvino]{V.A.~Victorov},
\author[ITEP]{V.E.~Vishnyakov},
\author[PNPI]{A.A.~Vorobyov},
\author[MPI]{K.~Vorwalter\thanksref{trq}},
\author[CMU,Fermi]{J.~You},
\author[Beijing]{Zhao~Wenheng},
\author[Beijing]{Zheng~Shuchen},
\author[Paulo]{R.~Zukanovich-Funchal}
\address[Ball]{Ball State University, Muncie, IN 47306, U.S.A.}
\address[Bogazici]{Bogazici University, Bebek 80815 Istanbul, Turkey}
\address[CMU]{Carnegie-Mellon University, Pittsburgh, PA 15213, U.S.A.}
\address[CBPF]{Centro Brasileiro de Pesquisas F\'{\i}sicas, Rio de Janeiro, Brazil}
\address[Fermi]{Fermi National Accelerator Laboratory, Batavia, IL 60510, U.S.A.}
\address[Protvino]{Institute for High Energy Physics, Protvino, Russia}
\address[Beijing]{Institute of High Energy Physics, Beijing, P.R. China}
\address[ITEP]{Institute of Theoretical and Experimental Physics, Moscow, Russia}
\address[MPI]{Max-Planck-Institut f\"ur Kernphysik, 69117 Heidelberg, Germany}
\address[MSU]{Moscow State University, Moscow, Russia}
\address[PNPI]{Petersburg Nuclear Physics Institute, St. Petersburg, Russia}
\address[Aviv]{Tel Aviv University, 69978 Ramat Aviv, Israel}
\address[SLP]{Universidad Aut\'onoma de San Luis Potos\'{\i}, San Luis Potos\'{\i}, Mexico}
\address[UFP]{Universidade Federal da Para\'{\i}ba, Para\'{\i}ba, Brazil}
\address[Bristo]{University of Bristol, Bristol BS8~1TL, United Kingdom}
\address[Iowa]{University of Iowa, Iowa City, IA 52242, U.S.A.}
\address[Flint]{University of Michigan-Flint, Flint, MI 48502, U.S.A.}
\address[Rome]{University of Rome ``La Sapienza'' and INFN, Rome, Italy}
\address[Paulo]{University of S\~ao Paulo, S\~ao Paulo, Brazil}
\address[Trieste]{University of Trieste and INFN, Trieste, Italy}
\thanks[tra]{deceased}
\thanks[trb]{Present address: Advanced Mask Technology Center, Dresden, Germany}
\thanks[trc]{Present address: University of California at Irvine, Irvine, CA 92697, USA}
\thanks[trd]{Present address: Instituto de F\'{\i}sica da Universidade Estadual de Campinas, UNICAMP, SP, Brazil}
\thanks[tre]{Present address: Kafkas University, Kars, Turkey}
\thanks[trf]{Present address: Physik-Department, Technische Universit\"at M\"unchen, 85748 Garching, Germany}
\thanks[trg]{Present address: The Boston Consulting Group, M\"unchen, Germany}
\thanks[trh]{Present address: Lucent Technologies, Naperville, IL}
\thanks[tri]{Present address: Baxter Healthcare, Round Lake IL}
\thanks[trj]{Present address: University of Alabama at Birmingham, Birmingham, AL 35294}
\thanks[trk]{Present address: NRCN, 84190 Beer-Sheva, Israel}
\thanks[trl]{Present address: S\"uleyman Demirel Universitesi, Isparta, Turkey}
\thanks[trm]{Present address: Legal Department, Oracle Corporation, Redwood Shores, California}
\thanks[trn]{Present address: DOE, Germantown, MD}
\thanks[tro]{Present address: Solidum, Ottawa, Ontario, Canada}
\thanks[trp]{ Present address: Siemens Medizintechnik, Erlangen, Germany}
\thanks[trq]{Present address: Allianz Insurance Group IT, M\"unchen, Germany}

\begin{abstract}
We report a precise measurement of the $\Omega_{c}^{0}$ lifetime.
The data were taken by the SELEX  (E781) experiment
using $600\,\mbox{GeV}/c$
$\Sigma ^{-}$, $\pi ^{-}$ and $p$ beams. 
The measurement has been made using $83\pm19$ reconstructed 
$\Omega ^{0}_{c}$ in the  $\Omega ^{-} \pi ^{-}\pi ^{+} \pi ^{+}$ and
$\Omega ^{-} \pi ^{+} $ decay modes. 
The lifetime of the  $\Omega_{c}^{0}$ is measured to be  
$65 \pm 13({\rm stat}) \pm 9 ({\rm sys})\,\mbox{fs}$.
\end{abstract}


\end{frontmatter}


\section{Introduction}

Several
experiments~\cite{Stiewe:1992hg,Frabetti:1994dp,Frabetti:1992bm,Cronin-Hennessy:2000bz,Frabetti:1995bi,Aubert:2006je}
in the last 
years have detected the $\Omega_{c}^{0}$ ground state
as well as an excited state.
Recently at Fermilab the photoproduction experiment, FOCUS, reported an 
observation  of a sample of 64~$\Omega_{c}^{0}$ events and they measured its 
lifetime as $72 \pm13 \pm11\,\mbox{fs}$~\cite{Link:2003nq}. 
The experiment WA89 published an $\Omega_{c}^{0}$ lifetime measurement of
$55^{+13~+18}_{-11~-23}\,\mbox{fs}$ from a sample of
86~events~\cite{Adamovich:1995pf}.
The present world average is  $69 \pm 12\,\mbox{fs}$ as reported in
Ref.~\cite{Yao:2006px}.  
Clearly additional measurements of lifetime as well as the branching ratios
with more statistical accuracy are needed to test theoretical
models~\cite{Bianco:2003vb,Cheng:1992gv}.

In this letter we report the results of a new measurement of the lifetime 
based on data from the hadroproduction experiment SELEX (E781) at Fermilab. 
The measurement is based on a sample of $83\pm19$ fully reconstructed 
$\Omega_{c} ^{0}$ from $15.3\times10^9$ hadronic interactions.

The SELEX detector at Fermilab is a 3-stage magnetic spectrometer.
The negatively charged $600\,\mbox{GeV}/c$
beam contains nearly equal fractions of 
$\Sigma$ and $\pi$. The positive beam contains $92\,\%$~protons.
Beam particles 
are identified as a baryon or a pion by a transition radiation detector.
The spectrometer was designed to study charm production in the forward 
hemisphere with good mass and decay vertex resolution for charm momenta
in the range $100-500\,\mbox{GeV}/c$.
Five interaction targets (2~Cu and 3~C) have a 
total target thickness of $4.2\,\%\,\lambda _{int}$ for protons.
The targets are
spaced by $1.5\,\mbox{cm}$.
Downstream of the targets there are 20~Silicon Strip 
Detectors (SSD) with a strip pitch of $20-25\,\mu\mbox{m}$
oriented in X,~Y,~U and
V views.  The first spectrometer level has three MultiWire Proportional 
Chambers (MWPC) with $3\,\mbox{mm}$ wire spacing and 
$2\times2\,\mbox{m}^2$ area downstream of 
bending magnet M1.  The second spectrometer level has 7~MWPCs with 
$2\,\mbox{mm}$ wire 
spacing downstream of the second bending magnet M2.  Each chamber has two 
sensitive planes in two orthogonal projections.  The scattered-particle 
spectrometers have momentum cutoffs of $2.5\,\mbox{GeV}/c$ 
and $15\,\mbox{GeV}/c$ respectively.
Typical momentum resolution for a $100\,\mbox{GeV}/c$ track is $0.5\,\%$.
A Ring-Imaging Cherenkov detector 
(RICH)~\cite{Engelfried:1998tv}, filled with Neon at room 
temperature and pressure, provides single track ring radius resolution of 
$1.4\,\%$ and $2\,\sigma$ $K/ \pi$ separation up to about $165\,\mbox{GeV}/c$.
A layout of the spectrometer can be found elsewhere~\cite{spec}.

\section{Reconstruction of hyperons {\mbox{\boldmath$\Omega $}},
{\mbox{\boldmath$\Xi$}} and {\mbox{\boldmath$\Sigma$}}}

The $\Omega _{c}^{0}$ decays studied here have a hyperon in the final state. 
Hyperons that decay upstream of or within the M1 magnet are called 
\emph{Kink} tracks. They are characterized by one charged track that decays 
to another charged particle and a neutral particle undetected by the 
spectrometer.  Such \emph{Kink} tracks differ from the majority of 
spectrometer tracks in that the vertex silicon track segment does not link to 
straight line tracks segments measured in the spectrometers after M1 and/or M2.
The \emph{Kink} reconstruction algorithm examines all unlinked Vertex SSD 
track segments that point to the M1 magnet aperture and tries to match each 
unlinked segment with an unlinked downstream track measured
in the M1/M2 spectrometer, using momentum-energy conservation with the\
hypothesis of a specific hyperon decay. 
The momentum of the parent hyperon ($\Omega^{\pm}$, $\Xi^{\pm}$ 
or $\Sigma^{\pm}$) is calculated using the assumed decay hypothesis.
The daughter $K^-$ in $\Omega^-$ decays must be RICH identified with the 
likelihood to be a $K$ exceeding that of its being a $\pi$.

\section{Data set and charm selection}

The charm trigger is very loose.  It requires a valid beam track, at least 
4~charged secondaries in the forward $150\,\mbox{mrad}$ cone,
and two hodoscope hits 
after the second bending magnet from tracks of charge opposite to that of the 
beam.  We triggered on about 1/3 of all inelastic interactions.
A computational filter linked MWPC tracks having momenta $>15\,\mbox{GeV}/c$ 
to hits in the vertex silicon and made a full reconstruction of these 
tracks together with a beam track to form primary and secondary vertices 
in the event. Events consistent with only a primary vertex are not saved. 
About 1/8  of all triggers are written to tape, for a final sample of about 
$10^9$~events.

In the full analysis the vertex reconstruction was repeated with tracks of all
momenta. The RICH detector identified charged tracks above $25\,\mbox{GeV}/c$.
Results reported here come from a second pass reconstruction through
the data, using a production code optimized for hyperon reconstruction. 

To separate the signal from the non-charm background we require that: 
(i)  the spatial separation $L$ between the reconstructed production and decay 
     vertices exceeds 6~times the combined error $\sigma_{L}$, 
(ii) each decay track, extrapolated to the primary vertex $z$ position, must 
     miss by a transverse distance $s\ge 2.5$ times its error $\sigma_{s}$, 
(iii) each candidate hyperon track, extrapolated to the \emph{Kink} vertex $z$ 
     position, must have a good vertex quality ($\chi^2/{\rm NDOF}<5$), 
(iv) the secondary vertex must lie outside any target material by at least 
     $0.05\,\mbox{cm}$, and 
(v)  decays must occur within a fiducial region.

\begin{figure}
\centering
\includegraphics[width=0.5\textwidth]{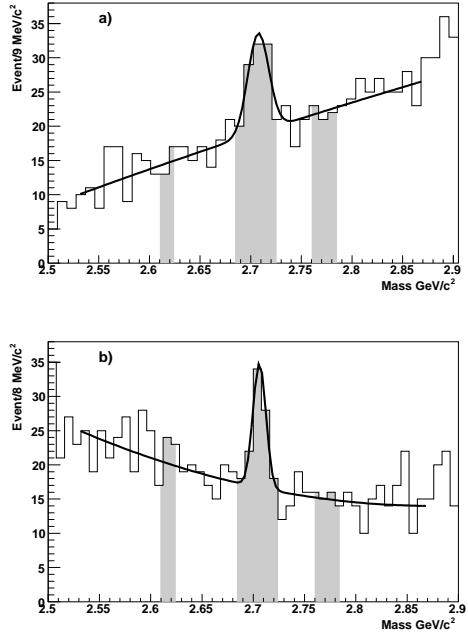}
\caption{ Invariant mass distribution for 
(a) $\Omega_{c}^{0} \rightarrow  \Omega^{-} \pi^{+} \pi^{-} \pi^{+}$, (b) $\Omega_{c}^{0} \rightarrow  \Omega^{-} \pi^{+} $. The shaded regions show the 
$\Omega_{c}^{0}$ signal and sideband regions. } 
\label{fig:mass}
\end{figure}

The total transverse momentum of pions from the
$ \Omega^{-} \pi^{+} \pi^{+}\pi ^{-}$ decay mode must be greater than 
$0.35\,\mbox{GeV}/c$ with respect to the $\Omega_{c}^{0}$ direction.
This cut optimizes the signal to background ratio.
We require a minimum $\pi$ momentum of $8\,\mbox{GeV}/c$ to reduce 
the number of fake invariant mass combinations. 
There are   $107\pm 22$  $\Omega_c^{0}$ candidates 
in three decay channels:  
$ \Omega^{-} \pi^{+} \pi^{+}\pi^{-}$, 
$ \Omega^{-} \pi^{+}$, and
$\Xi^{-} K^{-} \pi^{+} \pi^{+}$.
Details of the $\Omega_c^{0}$ mass measurement will be reported
elsewhere~\cite{sedat}.  
In the $\Xi^{-} K^{-} \pi^{+} \pi^{+}$ mode the signal is small and the signal 
to noise ratio is poor.  We've chosen not to include it in the lifetime 
measurement. The invariant mass distributions with tighter cuts for lifetime 
measurement are shown in fig.~\ref{fig:mass} 
for the two decay modes used here.  
 
\section{Lifetime evaluation using a maximum likelihood fit }

The average combined error $\sigma_{L}$ on the primary and secondary vertices 
and the average $\Omega _c ^{0}$ momentum give a proper time resolution of 
$16\,\mbox{fs}$, about $23\,\%$ of $\tau_{\Omega_c^0}$.
We used a binned maximum likelihood fitting technique with 
$5\,\mbox{fs}$ width bins to 
determine the $\Omega _c ^{0} $ lifetime. 
The fit was applied to a reduced proper time distribution, 
$ t^{*} ={M(L- L_{min})/ p c }$ where 
$M$ is the reconstructed charm mass, 
$p$ the reconstructed momentum, 
$L$ the measured vertex separation and
$L_{min}$ the minimum $L$ for each event to pass all the imposed selection 
cuts.  $L_{min}$ is determined event-by-event, along with the acceptance,
by the procedure described below.  We fitted all events with 
$t^{*} < 600\,\mbox{fs}$ in the mass range  
$ 2.685\,\mbox{GeV}/c^2 < M(\Omega _c ^{0}) < 2.725\,\mbox{GeV}/c^2$,
$ \pm ~2.5\,\sigma$ from the $\Omega _c ^{0} $ central mass value. 

The probability density function is:
\begin{eqnarray}
\nonumber
f(\tau _{\Omega _c ^{0}},\tau _{B1}, \tau _{B2} ,f_{B},f_{c},t^{*}) =\\
{(1-f_{B})N_{S}}{e^{-{t^{*}/{\tau _{\Omega _c ^{0}}}}}
\over{{\epsilon(t^{*}) \tau _{\Omega _c ^{0} }}}} + 
{f_{B}}N_{S} B(t^{*})
\label{eq:pdf}
\end{eqnarray}
where
\begin{equation}
B(t^{*})=f_{c}{{ e^{-{t^{*}/{\tau _{B1}}}}}\over{\tau _{B1}}} +
{(1-f_{c})}{{e^{-{t^{*}/{\tau _{B2}}}}}\over{\tau _{B2}}}
\end{equation}

The five parameters are: $\tau _{\Omega _c ^{0} }$ 
($ \Omega _c ^{0} $ lifetime), 
$\tau _{B1}$, $\tau _{B2}$ (background lifetimes), 
$f_{B}$ (background fraction in the
signal region) and $f_{c}$ (background splitting function). 
$N_{S}$ is the total number of events in the signal region.

The function is the sum of a term for the $\Omega _c ^{0} $
exponential decay corrected by the acceptance function $\epsilon (t^{*}) $
plus a background function $B(t^{*})$  consisting of two exponentials
to describe the strong decays and charm decays respectively. 
Its parameters were determined from the $t^{*}$ distribution from 
the $\Omega _c ^{0} $ sidebands.   
Together the mass widths of the sideband background windows,
 $2.610\,\mbox{GeV}/c^2 < M(\Omega _c ^{0} ) < 2.625\,\mbox{GeV}/c^2$   and
 $2.760\,\mbox{GeV}/c^2 < M( \Omega _c ^{0} ) < 2.785\,\mbox{GeV}/c^2$
was equal to the signal mass window.

The proper-time-dependent acceptance  $\epsilon (t^{*}) $ is independent 
of spectrometer features after the first magnet, e.g.,\ RICH efficiency and
tracking efficiency.  These efficiencies affect only the overall number of
events detected.  The proper time distribution of these events depends 
crucially on vertex reconstruction. To evaluate $\epsilon (t^{*}) $ we 
generated  $\Omega _c ^{0} $ events with a $(1-x_{F})^{3}$ distribution and 
decayed them using the QQ package~\cite{qq}.
We embedded these generated decays into real data events and reconstructed the 
embedded decays with the offline package including multiple Coulomb scattering 
in the spectrometer and the measured detector performance.
The correction function was evaluated as the fraction of the embedded events 
passing the selection cuts.  

\begin{figure}
\centering
\includegraphics*[width=0.5\textwidth]{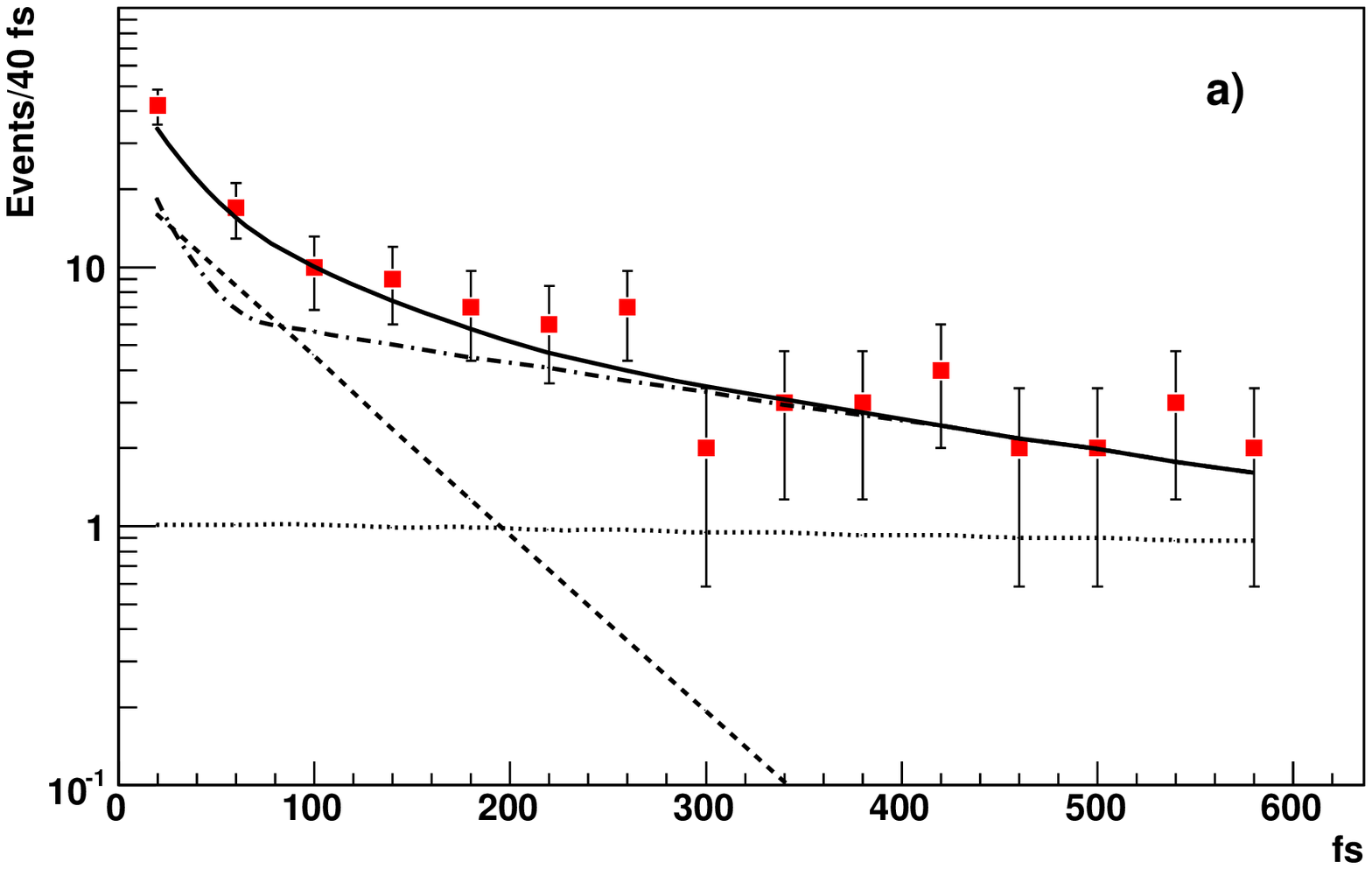}
\includegraphics*[width=0.5\textwidth]{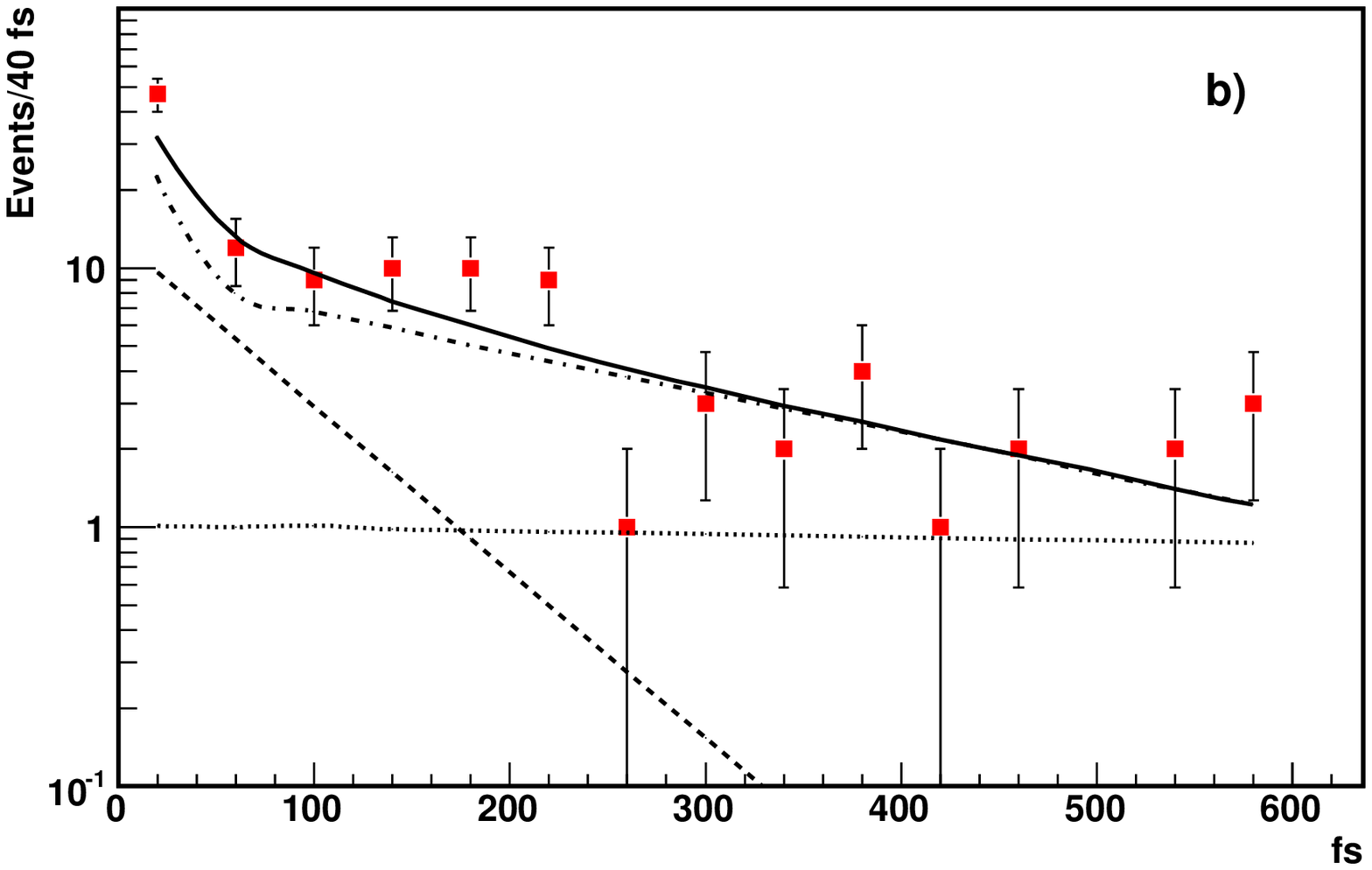}
\caption{Corrected reduced proper time distribution for events in the
$\Omega_c^{0}$ window $2706 \pm 20\,\mbox{MeV}/c^2$ (full boxes) and the 
results from the maximum likelihood fit (solid curve) for 
a) $ \Omega^{-} \pi^{+} \pi^{+}\pi^{-}$; b) $ \Omega^{-} \pi^{+}$.
The dashed curve shows the $\Omega_c^{0}$ proper lifetime, the dashed-dot
curve the fitted background and the dotted curve the acceptance.} 
\label{fig:life}
\end{figure}

Figure~\ref{fig:life} shows the overall fits to the data
distributions as a function of 
reduced proper time for  $\Omega^{-} \pi^{+} \pi^{-} \pi^{+}$ and  
$\Omega^{-} \pi^{+}$ decay modes. It also shows the acceptance function
$\epsilon (t^{*})$, which does not differ significantly from unity and is 
constant.  This is due to the fact  $L_{min}$ is chosen 6~times the combined 
error $\sigma_{L}$.  With SELEX's very high momentum and excellent resolution
this cut removes only the first $\sim1.5$ lifetimes from the sample.
\begin{table}
\centering
\caption{Lifetime fit results for the two $\Omega_c^0$ decay
modes analyzed. The errors are only statistical. The signal yields
from the fits to the mass plots in Fig.~\ref{fig:life} are also shown. }
\label{tab:fitresults}
\begin{tabular}{|l|c|c|}  \hline\hline
Fit results  &$\Omega^{-} \pi^{+} \pi^{+}\pi^{-}$&$\Omega^{-} \pi^{+}$\\
\hline
$\tau_{\Omega_{c}^{0}}$ $[\mbox{fs}]$ & $62.6 \pm 22.0 $  & $65.8 \pm 16.0$ \\
\hline
$\tau_{B1}$ $[\mbox{fs}]$ & $15.6 \pm 6.2 $  & $10.1 \pm 3.3 $  \\ \hline
$\tau_{B2}$ $[\mbox{fs}]$ & $388.2 \pm 27.0 $  & $281.4 \pm 22.5 $  \\ \hline
  Signal & $34 \pm 12 $  & $23 \pm 9 $  \\ \hline
  background & $84 \pm 13  $& $81 \pm 6 $\\ \hline\hline
  Signal yield  & $47 \pm 16  $& $36 \pm 11 $\\ \hline\hline
\end{tabular}
\end{table}

Table~\ref{tab:fitresults} summarizes the lifetime
fit results and the signal yields.  We measure 
an average lifetime $65 \pm13\,\mbox{fs}$.
The uncertainties are statistical only, 
evaluated where $-\ln{\mathcal{L}}$ increases by $0.5$.

\section{Systematic errors}

\begin{table} 
\centering
\caption{Systematic error contributions in fs for the two  
$\Omega_c^0$ decay modes analyzed.}
\label{tab:sig}
\begin{tabular}{|l|c|c|} \hline\hline
Source of uncertainty& 
$\Omega^{-} \pi^{+} \pi^{-} \pi^{+}$ , $\Omega^{-} \pi^{+} $   \\ \hline\hline
Vertex reconstruction & $<1\,\mbox{fs}$  \\ \hline
Acceptance function &  $1.75\,\mbox{fs}$ \\ \hline
Fit procedure  &  $8.5\,\mbox{fs}$   \\ \hline
Total systematic error & $8.6\,\mbox{fs}$  \\ \hline\hline
\end{tabular}
\end{table}

The systematic uncertainties for the $\Omega_c^{0}$ lifetime analysis are 
listed in Table~\ref{tab:sig} and described below.
Lifetime shifts due to reconstruction errors have been well studied in
our $D^0$ and $\Lambda_c$ work, with an order of magnitude higher
statistics~\cite{sasha,Kushnirenko:2000ed}.
Because of the high redundancy and good
precision of the silicon vertex detector, vertex mismeasurement effects are 
small at all momenta. Proper time assignment depends on correct momentum
determination. The SELEX momentum error is less than $0.5\,\%$
in all cases.  We 
assign a maximum systematic error from proper time
mismeasurement of $1\,\mbox{fs}$.
The acceptance function used in the fit was parameterized with a 1st and 2nd
order polynomial. The difference in lifetime result is $1.75\,\mbox{fs}$.
No significant difference in the lifetime correction function was found when we
changed the $n$ value of the $x_{F}$ distribution from 3 to 1. 
We varied the width of the sidebands and the bin size independently. 
The systematic error due to the fit procedure is $8.5\,\mbox{fs}$.

\section{Conclusions}

We have made a new measurement of the $\Omega _c ^{0}$ lifetime in
two independent decay channels,
$ \Omega^{-} \pi^{+} \pi^{+}\pi^{-}$, $ \Omega^{-} \pi^{+}$, using a maximum 
likelihood fit. SELEX measures the $\Omega _c ^{0}$ lifetime to be 
 $\tau_{ \Omega _c ^{0} } = 65 \pm 13({\rm stat}) \pm 9({\rm sys})\,\mbox{fs}$.
Our result is in excellent agreement with the world
average~\cite{Yao:2006px} and 
with the recent results published by the FOCUS
collaboration~\cite{Link:2003nq}.

\section*{Acknowledgment}
The authors are indebted to the staff of Fermi National Accelerator Laboratory
and for invaluable technical support from the staffs of collaborating
institutions.  
This project was supported in part by Bundesministerium f\"ur Bildung, 
Wissenschaft, Forschung und Technologie, Consejo Nacional de
Ciencia y Tecnolog\'{\i}a {(CONACyT)}, 
Conselho Nacional de Desenvolvimento Cient\'{\i}fico e Tecnol\'ogico,
Fon\-do de Apoyo a la Investigaci\'on (UASLP), 
Funda\c{c}\~ao de Amparo \`a Pesquisa do Estado de S\~ao Paulo (FAPESP), 
the Secretar\'{\i}a de Educaci\'on P\'ublica (Mexico) 
(grant number 2003-24-001-026), 
the Israel Science Foundation founded by the Israel Academy of Sciences and
Humanities, Istituto Nazionale di Fisica Nucleare (INFN), 
the International Science Foundation (ISF), 
the National Science Foundation (Phy \#9602178),
NATO (grant CR6.941058-1360/94), 
the Russian Academy of Science, 
the Russian Ministry of Science and Technology, 
the Russian Foundation for Basic Research: RFBR grant 05-02-17869,  
the Turkish Scientific and Technological Research Board (T\"{U}B\.ITAK),
the U.S. Department of Energy (DOE grant DE-FG02-91ER40664 and DOE contract
number DE-AC02-76CHO3000), and
the U.S.-Israel Binational Science Foundation (BSF).


\begin{thebibliography}{00}

\bibitem{Stiewe:1992hg}
  J.~Stiewe,
  AIP Conf.\ Proc.\  {\bf 272} (1993) 1076.

\bibitem{Frabetti:1994dp}
  P.~L.~Frabetti {\it et al.}  [E687 Collaboration],
  Phys.\ Lett.\  B {\bf 338} (1994) 106.

\bibitem{Frabetti:1992bm}
  P.~L.~Frabetti {\it et al.}  [E687 Collaboration],
  Phys.\ Lett.\  B {\bf 300} (1993) 190.


\bibitem{Cronin-Hennessy:2000bz}
  D.~Cronin-Hennessy {\it et al.}  [CLEO Collaboration],
  Phys.\ Rev.\ Lett.\  {\bf 86} (2001) 3730
  [arXiv:hep-ex/0010035].

\bibitem{Frabetti:1995bi}
  P.~L.~Frabetti {\it et al.}  [E687 Collaboration],
  Phys.\ Lett.\  B {\bf 357} (1995) 678.

\bibitem{Aubert:2006je}
  B.~Aubert {\it et al.}  [BABAR Collaboration],
  Phys.\ Rev.\ Lett.\  {\bf 97} (2006) 232001
  [arXiv:hep-ex/0608055].

\bibitem{Link:2003nq}
  J.~M.~Link {\it et al.}  [FOCUS Collaboration],
  Phys.\ Lett.\  B {\bf 561} (2003) 41
  [arXiv:hep-ex/0302033].

\bibitem{Adamovich:1995pf}
  M.~I.~Adamovich {\it et al.}  [WA89 Collaboration],
  Phys.\ Lett.\  B {\bf 358} (1995) 151
  [arXiv:hep-ex/9507004].

\bibitem{Yao:2006px}
  W.~M.~Yao {\it et al.}  [Particle Data Group],
  J.\ Phys.\ G {\bf 33} (2006) 1.

\bibitem{Bianco:2003vb}
  S.~Bianco, F.~L.~Fabbri, D.~Benson and I.~Bigi,
  Riv.\ Nuovo Cim.\  {\bf 26N7} (2003) 1
  [arXiv:hep-ex/0309021].

\bibitem{Cheng:1992gv}
  H.~Y.~Cheng,
  Phys.\ Lett.\  B {\bf 289} (1992) 455.

\bibitem{Engelfried:1998tv}
  J.~Engelfried {\it et al.}  [SELEX Collaboration],
  Nucl.\ Instrum.\ Meth.\  A {\bf 431} (1999) 53
  [arXiv:hep-ex/9811001].

\bibitem{spec}   J.S. Russ {\it et al.} [SELEX Collaboration], 
                  in \emph{ Proceedings of the
                 29th International Conference on High Energy Physics,}
                 1998, edited by A. Astbury \emph{et al.} (Word Scientific,
                 Singapore, 1998) Vol. II, p. 1259 [arXiv:hep-ex/9812031].

\bibitem{sedat}  SELEX Collaboration, to be published.

\bibitem{qq}     {\tt http://www.lns.cornell.edu/public/CLEO/soft/QQ/}

\bibitem{sasha}  A.Y. Kushnirenko, Ph.D.\ Thesis, Carnegie Mellon
                 University, 2000 (unpublished). FERMILAB-THESIS-2000-09.

\bibitem{Kushnirenko:2000ed}
  A.~Kushnirenko {\it et al.}  [SELEX Collaboration],
  Phys.\ Rev.\ Lett.\  {\bf 86} (2001) 5243
  [arXiv:hep-ex/0010014].

\end{thebibliography}
\end{document}